\documentclass[prd,twocolumn,aps,showpacs,floats]{revtex4}
\usepackage{graphicx}
\usepackage{bm}
\usepackage{epsfig}
\usepackage{amsmath,amsfonts}
\usepackage{url}

\begin{document}

\title{Exact cosmological solutions from Hojman conservation quantities}

\author{Salvatore Capozziello$^{1.2}$}\email{capozzie@na.infn.it}
\author{ Mahmood Roshan$^3$}\email{rowshan@alumni.ut.ac.ir}
\affiliation{$^1$Dipartimento di  Fisica, Universit\`{a} di Napoli "Federico II", Napoli, Italy}
\affiliation {$^2$INFN Sez. di Napoli, Compl. Univ. di Monte S. Angelo, Edificio G, Via Cinthia, I-80126, Napoli, Italy}
\affiliation{$^3$Department of Physics, Ferdowsi University of Mashhad, P.O. Box 1436, Mashhad, Iran}

\begin{abstract}
We present a new approach to find exact solutions for cosmological models. By requiring the existence of a symmetry transformation vector for the equations of motion of the given cosmological model (without using either Lagrangian or Hamiltonian), one can find corresponding Hojman conserved quantities. With the help of these conserved quantities, the analysis of the cosmological model can be simplified. In the case of quintessence scalar-tensor models, we show that the Hojman conserved quantities exist for a wide range of   $V(\phi)$-potentials and allow to find  exact solutions for the cosmic scale factor and the scalar field. Finally, we investigate the general cosmological behavior of solutions by adopting a phase-space view.
\end{abstract}


\date{\today}

\maketitle

\section{Introduction}
Symmetries are a fundamental tool to study  physical systems since they allow to reduce dynamics  and give insight into  conserved quantities. For example, the Noether  theorem has been widely adopted in classical and quantum physics as well as  in general relativity and cosmology. As an example in general relativity, it is well-known that the black hole entropy is a Noether conserved quantity \cite{wald}. In cosmology, the so called \emph{Noether symmetry approach} provides a method to solve exactly  dynamics \cite{ritis}. This approach is based on requiring the existence of a Noether symmetry for the point-Lagrangian associated to the equations of motion for  a given cosmological model. It has been used in several alternative theories of gravity \cite{alternative} in order to find out exact cosmological solutions, and also to fix the form of undetermined couplings  and potentials  \cite{noether}.

In this paper, we show that Hojman conservation theorem \cite{hojman} can also provide an approach to find exact solutions for the cosmological models. As we shall see below, there is a main difference between the Noether conservation theorem and the Hojman conservation theorem. The first  relies on  the Lagrangian (and Hamiltonian) structure of the equations of motion. In the Hojman picture,  there is no need for  Lagrangians and Hamiltonian functions given a priori. The symmetry vectors and the corresponding conserved quantities can be obtained by using directly the equations of motion. Specifically, unlike the Noether symmetry approach, it is not needed to find first a point-Lagrangian for the equations of motion. Moreover, these two approaches may give rise to  different conserved quantities. For example, in the case of minimally coupled scalar-tensor gravity models, the Noether symmetry exists only for exponential potential $V(\phi)$ \cite{ritis}; however, as we shall see below, the Hojman symmetry exists for a wide range of potentials. In principle, the Hojman conserved quantities  theorem could be a different class with respect to the Noether ones because they do not require that  field equations  directly derive  from a Lagrangian.

In particular, using the Hojman approach, we find a new class of scalar-field potentials. By a  dynamical system analysis, these model can be investigated  for some  simple cases. We will show that there exist attractor solutions corresponding to late time scalar-field dominated universe and then interesting to address the dark energy problem.

The layout of the paper is the  following. In Sec. \ref{hojman ct}, we briefly review the Hojman conservation theorem. Secs: \ref{HFE} and \ref{quintessence} are devoted to the application of  the Hojman conservation theorem to given scalar-tensor models where we find out exact solutions. In Sec. \ref{newpot}, we investigate the  relevant phase-space features of the cosmological models. Conclusions are drawn in Sec. \ref{concl}.

\section{The Hojman conservation theorem}\label{hojman ct}
In order to point out the differences between the Hojman and the Noether approaches, let us derive the conservation laws  in the first case.
Consider a set of second-order differential equations
\begin{equation}
\ddot{q}^{i}=F^{i}(q^j,\dot{q}^j,t)\ \ \ \ \ \ \  i,j=1,...,n
\label{neq}
\end{equation}
where  dots indicate derivatives with respect to time, namely "$~\dot{}~"=d/dt$. In the case of cosmology, it is  appropriate to interpret the dot as the differentiation with respect to  the cosmic time.
Before stating the Hojman conservation theorem,  let us recall the definition of the symmetry vector for the above equations. The symmetry vector $X^{i}$ is defined so that the infinitesimal transformation
\begin{equation}
q'^{i}=q^i+\epsilon X^i(q^j,\dot{q}^j,t)
\end{equation}
maps solutions $q^i$ of the equations \eqref{neq} into solutions $q'^i$ of the same equation \cite{lutzky}. It is straightforward to verify that the symmetry vector for equations \eqref{neq} satisfies
\begin{equation}\label{xe}
\frac{d^2 X^i}{dt^2}-\frac{\partial F^i}{\partial q^j}X^j-\frac{\partial F^i}{\partial \dot{q}^j}\frac{dX^j}{dt}=0
\end{equation}
with
\begin{equation}
\frac{d}{dt}=\frac{\partial}{\partial t}+\dot{q}^i\frac{\partial}{\partial q^i}+F^i\frac{\partial}{\partial \dot{q}^i}.
\end{equation}
Keeping in mind the definition of the symmetry vector $X^i$, the Hojman conservation quantities  can be defined by the  following \cite{hojman}
\\
\\
\textbf{Theorem:}
\\
\\
\emph{
1. if the "force" $F$ satisfies}
\begin{equation}
\frac{\partial F^i}{\partial \dot{q}^i}=0
\end{equation}
\emph{ then}
\begin{equation}\label{cq1}
Q=\frac{\partial X^i}{\partial q^i}+\frac{\partial}{\partial \dot{q}^i}\left(\frac{dX^i}{dt}\right)
\end{equation}
\emph{is a conserved quantity for equations \eqref{neq}, i.e.  $\frac{dQ}{dt}=0$.}
\\
\\
\emph{2. and if $F$ satisfies}
\begin{equation}\label{lambda}
\frac{\partial F^i}{\partial \dot{q}^i}=-\frac{d}{dt}\ln\gamma
\end{equation}
\emph{where $\gamma$ is a function of $q^i$, then}
\begin{equation}\label{cq2}
Q=\frac{1}{\gamma}\frac{\partial(\gamma  X^i)}{\partial q^i}+\frac{\partial}{\partial \dot{q}^i}\left(\frac{dX^i}{dt}\right)
\end{equation}
\emph{is a conserved quantity.}\\
Such a theorem can be applied to dynamical systems describing cosmological models as we will show below.

\section{Hojman conserved quantities for cosmological  equations}\label{HFE}
As an example for the above considerations, let us consider  the case of spatially flat Friedmann-Robertson-Walker (FRW) space-time.  The Friedman equations are
\begin{equation}\label{fre2}
\ddot{a}=-\frac{3p+\rho}{6}a
\end{equation}
\begin{equation}\label{fre1}
\dot{a}^2=\frac{\rho}{3}a^2
\end{equation}
where $a(t)$ is the cosmic scale factor, $p$ is the pressure of the cosmic fluid and $\rho$ is the corresponding energy density (we work in units where $c=8\pi G=1$).  If we assume that the cosmic perfect fluid obeys the  equation of state
\begin{equation}
p=w\rho
\end{equation}
where $-1<w<1$, then combining equations \eqref{fre2} and \eqref{fre1} we get
\begin{equation}\label{fre3}
\ddot{a}=\alpha \frac{\dot{a}^2}{a}
\end{equation}
where $\alpha=-\frac{3w+1}{2}$. Mathematically, equation \eqref{fre3} can be considered as a one dimensional "motion" under the action of a "position" and "velocity" dependent "force". Comparing this equation with \eqref{neq}, it is clear that $F(a,\dot{a})=\alpha \frac{\dot{a}^2}{a}$. Thus one can easily show that
\begin{equation}\label{lam}
\gamma(a)=a^{-2\alpha}
\end{equation}
For the sake of simplicity, let us assume that the one dimensional symmetry vector $X$ dose not depend explicitly on $t$, i.e. $X=X(a,\dot{a})$. In this case, equation \eqref{xe} takes the form
\begin{equation}\label{sv1}
\begin{split}
&\left(-\frac{\alpha}{a}\frac{\partial X}{\partial a}+\frac{\partial^2 X}{\partial a^2}+\frac{\alpha}{a^2}X\right)+\dot{a}^2\frac{\alpha^2}{a^2}\frac{\partial^2 X}{\partial \dot{a}^2}+ \\& \dot{a}\left(\frac{2\alpha}{a}\frac{\partial^2 X}{\partial a\partial \dot{a}}-\frac{\alpha}{a^2}\frac{\partial X}{\partial \dot{a}}\right)=0
\end{split}
\end{equation}
It is easy to verify that $X(a,\dot{a})\sim a^n \dot{a}^m$ is a solution for this equation provided that $n=1-m \alpha$. For this symmetry vector, using equations \eqref{cq2} and \eqref{lam}, we find the following conserved quantity
\begin{equation}\label{cq3}
Q_0=(m+2)(\alpha-1)a^{n-1}\dot{a}^m
\end{equation}
Assuming $w\neq-1$, we can rewrite \eqref{cq3} as
\begin{equation}\label{cheg}
a^{\frac{1+3w}{2}}\dot{a}=c_1=const
\end{equation}
This nonlinear differential equation can be easily solved, giving
\begin{equation}
    a(t)=\left[\frac{3(1+w)}{2}(c_1 t+c_2)\right]^{\frac{2}{3(1+w)}}
\end{equation}
where $c_2$ is an integration constant. In the case of dark energy density ($w=-1$) $Q_0$ vanishes, and equation \eqref{cq3} does not lead to an exact solution for the cosmic scale factor.

 It should be stressed that, using equation \eqref{fre1}, we can rewrite equation \eqref{cheg} as $\rho a^{3(1+w)}=const$. Nevertheless, even without working through the details of finding the symmetry vector, this expression can be directly derived from the Friedmann equations. However, as we shall see in the next section, in the context of other cosmological models, where there exist some undetermined functions in the field equations, finding the Hojman conserved quantity is not possible without calculating the symmetry vector.
\section{Hojman conserved quantities and scalar-tensor cosmology}
\label{quintessence}
A time dependent vacuum energy is sometimes called \emph{quintessence}, and  it is described by a scalar field $\phi$ minimally coupled to gravity \cite{ratra}. The potential of the scalar field $V(\phi)$ is chosen in a way that leads to late time accelerated expansion. In a flat FRW space-time, the equations of motion are given by
\begin{equation}\label{quin1}
\frac{\dot{a}^2}{a^2}=\frac{1}{3}\left[\frac{\dot{\phi}^2}{2}+V(\phi)\right]
\end{equation}
\begin{equation}\label{quin2}
\ddot{a}=\frac{a}{3}\left[V(\phi)-\dot{\phi}^2\right]
\end{equation}
We recall that the time derivative of \eqref{quin1} together with \eqref{quin2} yield
\begin{equation}\label{quin3}
\ddot{\phi}+3\frac{\dot{a}}{a}\dot{\phi}+V'(\phi)=0
\end{equation}
which is a Klein-Gordon equation,  the equation of motion of the scalar field. Hereafter the prime means $h'(y)=dh(y)/dy$. By introducing the variable $x=\ln a$, and by combining equations \eqref{quin1} and \eqref{quin2}, we find the following equation of motion
\begin{equation}\label{qem}
\ddot{x}=-f(x) \dot{x}^2
\end{equation}
where
\begin{equation}\label{f}
f(x)=\frac{1}{2} \phi'(x)^2
\end{equation}
With the assumption that  $a(t)$ and $\phi(t)$ are invertible functions of $t$, the dynamics governed by equations \eqref{quin1}-\eqref{quin3} can be reduced to a one dimensional motion. In this case, equation \eqref{quin3} can be considered as a constraint equation. From equation \eqref{qem}, it is clear that $F(x,\dot{x})=-f(x)\dot{x}^2$, thus
\begin{equation}\label{quinl0}
\gamma(x)=\gamma_0 e^{\int f(x) dx}
\end{equation}
where $\gamma_0$ is an integration constant. Another useful differential equation can be obtained by dividing equations \eqref{quin3} and \eqref{quin1}. The result is
\begin{equation}\label{pot1}
\frac{V'(\phi)}{V(\phi)}=\frac{f(x)\phi'(x)-\phi''(x)-3\phi'(x)}{3-\frac{1}{2}\phi'(x)^2}
\end{equation}
If we assume that the one dimensional symmetry vector $X$ does not depend explicitly on time, then equation \eqref{xe} takes the following form
\begin{equation}\label{quin11}
\begin{split}
&\left(f(x)\frac{\partial X}{\partial x}+f'(x)X+\frac{\partial^2 X}{\partial x^2}\right)+\dot{x}^2 f(x)^2 \frac{\partial^2 X}{\partial \dot{x}^2}-\\&\dot{x}\left(2f(x)\frac{\partial^2 X}{\partial x\partial \dot{x}}+f'(x)\frac{\partial X}{\partial \dot{x}}\right)=0
\end{split}
\end{equation}
In order to find some solutions for this equation, we consider the following cases:

\subsection{The case  $X=X(\dot{x})$}

In this case, the only solution for \eqref{quin11} is
\begin{equation}\label{x1}
X(\dot{x})=A_0 \dot{x}^n+A_1 \dot{x}
\end{equation}
and
\begin{equation}
f(x)=-\left(\frac{1}{nx+f_0}\right)
\end{equation}
where $A_0, A_1, n$ and $f_0$ are constant parameters. It is straightforward to verify that the symmetry vector $X\sim \dot{x}$ leads to a zero constant of motion ($Q=0$). Therefore, without any loose of generality, we set $A_1=0$ and $n\neq1$. Let us define a new variable as $y=-(nx+f_0)>0$. Note that $f(x)$ is a positive function. Now, by introducing $f(y)=\frac{1}{y}$ into equation \eqref{f}, we find
\begin{equation}\label{fifi}
\phi(y)=\phi_c\pm\frac{1}{n}\sqrt{8y}
\end{equation}
where $\phi_c$ is a constant. Furthermore, we can rewrite equation \eqref{pot1} as
\begin{equation}
\frac{V'(y)}{V(y)}=\frac{2}{ny}-\frac{1}{y(1-3y)}
\end{equation}
The solution of this first order differential equation is given by
\begin{equation}
V(y)=V_0 y^{\frac{2-n}{n}}(3y-1)
\end{equation}
Using equation \eqref{fifi}, the generic potential with respect to $\varphi=\phi-\phi_c$ is
\begin{equation}\label{pot2}
V(\varphi)=\lambda \varphi^{\frac{4}{n}}-\frac{8\lambda}{3n^2}\varphi^{\frac{4}{n}-2}
\end{equation}
where
\begin{equation}
\lambda=3V_0\left(\frac{n^2}{8}\right)^{\frac{2}{n}}
\end{equation}
For $n<0$, the potential \eqref{pot2} gives rise to the so called \emph{intermediate inflationary model} \cite{barrow,muslimov}. In the case of $n=1$, this potential leads to the so called $\lambda \varphi^{4}$ model \cite{linde}. However, note that for $\lambda \varphi^{4}$ model there is no symmetry vector in the form presented in \eqref{x1}.

On the other hand, using equations \eqref{cq2} and \eqref{quinl0} we find the Hojman conserved quantity as follows
\begin{equation}\label{cq10}
\frac{\dot{x}^n}{nx+f_0}=Q_0
\end{equation}
if $\dot{x}^n>0$ then $Q_0<0$, and if $\dot{x}^n<0$ then $Q_0>0$. Also, since $\dot{x}$ can be negative or positive,  we assume that $n$ is an integer number. Thus we  rewrite \eqref{cq10} as
\begin{equation}\label{cq11}
\dot{y}^n=(-n)^n y |Q_0|
\end{equation}
The solution of this equation is given by
\begin{equation}\label{cq12}
y(t)=\left[(1-\frac{1}{n})(y_0-n|Q_0|^{\frac{1}{n}}t)\right]^{\frac{n}{n-1}}
\end{equation}
where $y_0$ is an integration constant. Consequently, using equations \eqref{fifi} and \eqref{cq12}, we can summarize the exact solution of $a(t)$ and $\varphi(t)$ for the potential \eqref{pot1} as follows
\begin{equation}\label{exact1}
\begin{split}
&a(\tau)=e^{-\frac{f_0}{n}}e^{-\frac{1}{n}\left[(1-\frac{1}{n})\tau\right]^{\frac{n}{n-1}}}\\&
\varphi(\tau)=\pm \frac{\sqrt{8}}{n}\left[(1-\frac{1}{n})\tau\right]^{\frac{n}{2(n-1)}}
\end{split}
\end{equation}
where the parameter $\tau$ is defined as
\begin{equation}
\tau=y_0-n|Q_0|^{\frac{1}{n}}t
\end{equation}
Substituting these solutions together with the potential \eqref{pot2} into equation \eqref{quin3}, we get
\begin{equation}
V_0=|Q_0|^{\frac{2}{n}}
\end{equation}
Thus, a symmetry vector $X\sim \dot{x}^n$ exists for equation \eqref{qem}, if the generic potential $V(\varphi)$ takes the form  \eqref{pot2}. Now, let us consider more general situations where the symmetry vector $X$ is a function of both $x$ and $\dot{x}$.
\subsection{The case $X(x,\dot{x})=\dot{x} g(x)$}
Here $g(x)$ is an arbitrary function of $x$, which we call \emph{generating function}. In this case, using equation \eqref{quin11}, one can show that $X$ is  a symmetry vector if the generating function satisfies
\begin{equation}\label{ex21}
f(x)=\frac{1}{2}\phi'(x)^2=\frac{g''(x)}{g'(x)}>0
\end{equation}
Solving this equation, we find the scalar field $\phi$ with respect to $x$, namely
\begin{equation}\label{ex22}
\phi(x)=\phi_c\pm \sqrt{2}\int \sqrt{\frac{g''(x)}{g'(x)}}dx
\end{equation}
On the other hand, the corresponding Hojman conserved quantity is given by
\begin{equation}\label{ex23}
\dot{x}g'(x)=Q_0
\end{equation}
In general, this is a nonlinear first order differential equation for $x(t)$ and can lead to an exact solution for the cosmic scale factor.

By using equations \eqref{pot1} and \eqref{ex21}, one can find the generic potential with respect to $x$
\begin{equation}\label{ex24}
V(x)=V_0\frac{g''(x)-3g'(x)}{g'(x)^3}
\end{equation}
It is obvious that the potential can be expressed in terms of the first and second derivatives of the generating function. Substituting this potential into \eqref{quin3} and using \eqref{ex23}, we find $V_0=-Q_0^2$. Thus the potential $V(x)$ can be rewritten as
\begin{equation}\label{ex25}
V(x)=Q_0^2\frac{3g'(x)-g''(x)}{g'(x)^3}=Q_0^2(3-f(x))e^{-2\int f(x)dx}
\end{equation}
In the following, we choose some specific forms for the generating function and find some exact solutions.

\subsubsection {$g(x)=\lambda e^{\frac{\alpha^2}{2}x}$}
Where $\alpha$ and $\lambda$ are constant parameters. In this case $f(x)=\frac{\alpha^2}{2}>0$. Using equation \eqref{ex22} we get
\begin{equation}\label{ex26}
\phi(x)=\phi_0\pm \alpha (x-x_0)
\end{equation}
where $\phi_0$ and $x_0$ are the values of $\phi(t)$ and $x(t)$ at $t=t_0$. Also, the Hojman conserved quantity \eqref{ex23} yields to the following solution for $x(t)$
\begin{equation}\label{ex27}
x(t)=x_0+\frac{2}{\alpha^2}\ln \left[1+\frac{Q_0 \alpha^2}{2\lambda}e^{-\frac{\alpha^2}{2}x_0}(t-t_0)\right]
\end{equation}
Furthermore, substituting the expression $f(x)=\frac{\alpha^2}{2}$ into \eqref{ex25} and using \eqref{ex26}, we find
\begin{equation}\label{ex28}
V(\varphi)=\frac{2(6-\alpha^2)Q_0^2}{\alpha^4 \lambda^2}e^{\mp\alpha \varphi}
\end{equation}
where $\varphi=\phi-\phi_0$. We summarize the solutions for $a(t)$ and $\varphi(t)$ for the above generic potential as follows
\begin{equation}\label{ex29}
\begin{split}
&a(t)=e^{x_0}\left[1+\frac{Q_0 \alpha^2}{2\lambda}e^{-\frac{\alpha^2}{2}x_0}(t-t_0)\right]^{\frac{2}{\alpha^2}}\\&
\varphi(t)=\pm \frac{2}{\alpha}\ln \left[1+\frac{Q_0 \alpha^2}{2\lambda}e^{-\frac{\alpha^2}{2}x_0}(t-t_0)\right]
\end{split}
\end{equation}

\subsubsection{$g(x)=\frac{(f_0+x)^{1+\alpha}}{1+\alpha}~~~~~(\alpha>0)$}
Where $\alpha$ and $f_0$ are constant parameters. In this case, $f(x)=\frac{\alpha}{f_0+x}$. Thus $f(x)>0$ if $f_0+x>0$. Using equation \eqref{ex22} we find
\begin{equation}\label{ex31}
\phi(x)=\phi_0\pm \sqrt{8\alpha (f_0+x)}\mp\sqrt{8\alpha (f_0+x_0)}
\end{equation}
Furthermore, the Hojman conserved quantity \eqref{ex23} leads to
\begin{equation}\label{ex32}
x(t)=-f_0+\left((1+\alpha)Q_0(t-t_0)+(f_0+x_0)^{1+\alpha}\right)^{\frac{1}{1+\alpha}}
\end{equation}
And the generic potential $V(\varphi)$ is given by
\begin{equation}\label{ex33}
V(\varphi)=\lambda \varphi^{-4\alpha} -\frac{8}{3}\lambda\alpha^2\varphi^{-4\alpha-2}
\end{equation}
where $\varphi=\phi-\phi_0\pm\sqrt{8\alpha (f_0+x_0)}$ and $\lambda=3Q_0^2(64\alpha^2)^{\alpha}$. Note that if $\alpha=-\frac{1}{n}$ ($n<0$), the above potential coincides with potential \eqref{pot2}. This shows that different symmetry vectors can lead to the  same potential.
 However, considering that $\alpha$ is a positive real number and $n$ is an integer number, potentials \eqref{ex33} is more general than \eqref{pot2}.

 \subsubsection{$g(x)=-\sqrt{2}\alpha\left(x+e^{2\alpha^2 x}\right)$}
 Where $\alpha$ is a real number. In this case, one can easily verify that $f(x)>0$ for every $\alpha$. Using the Hojman conserved quantity, we have
 \begin{equation}\label{ex41}
x(t)=\frac{1}{2\alpha^2}\left[\tau-W(2\alpha^2e^{\tau})\right]
\end{equation}
where $W$ is the \emph{Lambert W function} and $\tau$ is defined as
 \begin{equation}
\tau=-\sqrt{2}\alpha(g(x_0)+Q_0(t-t_0))
\end{equation}
Note that as before $x_0=x(t_0)$. Also, the scalar field $\phi$ with respect to $x$ is given by
 \begin{equation}\label{ex42}
\phi(x)=\phi_0\pm \frac{2}{\alpha}\textrm{arcsinh}\left(\sqrt{2}\alpha e^{\alpha^2 x}\right)\mp\frac{2}{\alpha}\textrm{arcsinh}\left(\sqrt{2}\alpha e^{\alpha^2 x0}\right)
\end{equation}
Finally, the generic potential with respect to $\varphi$ can be written as
\begin{equation}\label{pot4}
V(\varphi)=Q_0^2\left[\textrm{sech}^6\left(\frac{\alpha\varphi}{2}\right)-\frac{2\alpha^2-3}{2\alpha^2}\textrm{sech}^4\left(\frac{\alpha\varphi}{2}\right) \right]
\end{equation}
Where $\varphi=\phi-\phi_0\pm\frac{2}{\alpha}\textrm{arcsinh}\left(\sqrt{2}\alpha e^{\alpha^2 x0}\right)$.
Now let us summarize the exact solution for the above potential
\begin{equation}\label{ex43}
\begin{split}
&a(\tau)=e^{\frac{1}{2\alpha^2}\left[\tau-W(2\alpha^2e^{\tau})\right]}\\&
\varphi(\tau)=\frac{2}{\alpha}\textrm{arcsinh}\left[\sqrt{2}\alpha~e^{\frac{1}{2\alpha^2}\left(\tau-W(2\alpha^2e^{\tau})\right)}\right]
\end{split}
\end{equation}

In this model, if $\alpha^2\leq \frac{3}{2}$,  $V(\varphi)>0$ and so it can be considered as a quintessence model. We will consider some important features of such a  model in Sec. \ref{newpot}.

\subsubsection{$g(x)=\frac{1}{\alpha}\ln \left[\tanh\left(\frac{\alpha\phi(x)}{2}\right)\right]$}
\label{next}
If the generating function is given as a function of $\phi$, then the equation \eqref{ex21} can be rewritten as follows
\begin{equation}\label{ex51}
\left(\frac{dx}{d\phi}\right)^{-1}\left(\frac{1}{2}+\frac{d^2x}{d\phi^2}\right)=\frac{g''(\phi)}{g'(\phi)}
\end{equation}
This is a nonlinear differential equation for $x(\phi)$. On the other hand, the Hojman conserved quantity can be written as
\begin{equation}\label{ex52}
\dot{\phi}g'(\phi)=Q_0
\end{equation}
Furthermore, using equation \eqref{pot1} the generic potential can be written directly with respect to $\phi$
\begin{equation}\label{ex53}
V(\phi)=Q_0^2(3-f(\phi))e^{\mp\int \sqrt{2f(\phi)}d\phi}
\end{equation}
Now, let us solve equation \eqref{ex51} for the given generating function $g(\phi)$. The result is
\begin{equation}\label{ex54}
x(\phi)=\frac{\beta}{2\alpha^2}\ln\left[\tanh\left(\frac{\alpha\phi}{2}\right)\right]-\frac{1}{2\alpha^2}\ln \left[\sinh (\alpha\phi)\right]
\end{equation}
where $\beta$ is an integration constant. Thus $f(\phi)$ is
\begin{equation}\label{ex55}
f(\phi)=\frac{2\alpha^2\textrm{sinh}^2(\alpha\phi)}{[\beta-\cosh(\alpha\phi)]^2}>0
\end{equation}
With $f(\phi)$ in hand, we can find the potential $V(\phi)$ using equation \eqref{ex53}, namely
\begin{equation}\label{ex56}
\begin{split}
&V(\phi)=Q_0^2(3-2\alpha^2)\left[\cosh(\alpha\phi)-\frac{\beta}{1-\frac{2}{3}\alpha^2}\right]^2\\&~~~~~~~~~~~~~~
+2\alpha^2Q_0^2\left(\frac{\beta^2}{1-\frac{2}{3}\alpha^2}\right)
\end{split}
\end{equation}
If we assume that $\beta=1-\frac{2}{3}\alpha^2$ then the potential \eqref{ex56} can be rewritten as
\begin{equation}\label{ex570}
V(\phi)=V_0 [\cosh(\alpha\phi)-1]^2+\Lambda
\end{equation}
where $V_0=3Q_0^2\beta$ and $\Lambda=3Q_0^2(1-\beta)\beta$. This potential is similar to the potential suggested by Sahni and Wang \cite{sahni}. We recall that the generic potential of the quintessence model introduced in \cite{sahni}, is given by
\begin{equation}\label{ex5700}
V(\phi)=V_0 [\cosh(\alpha\phi)-1]^p
\end{equation}
where $\alpha$ and $p$ are constant parameters. Although the parameter $\Lambda$ is zero in \cite{sahni} (note that $\Lambda$ can not be zero in the potential \eqref{ex570}), the potential \eqref{ex570} is similar to the potential \eqref{ex5700} with $p=2$.

Equation \eqref{ex52} yields to the following solution for $\phi(t)$
\begin{equation}\label{ex57}
\phi(\tau)=\frac{2}{\alpha}\textrm{arctanh}(e^{\tau})
\end{equation}
in which $\tau=\alpha(Q_0 t+\phi_c)$, and $\phi_c$ is an integration constant. Substituting this solution into \eqref{ex54}, the cosmic scale factor is
\begin{equation}\label{ex58}
a(\tau)=e^{\frac{\beta}{2\alpha^2}\tau}\left[\sinh(2\textrm{arctanh}(e^{\tau}))\right]^{-\frac{1}{2\alpha^2}}
\end{equation}
As final remark, it is important to stress again that  these exact solutions strictly depend on the Hojman conserved quantities that allow to reduce the dynamical system.

\section{The space-phase view of the model}
\label{newpot}
\begin{figure}[!t]
\hspace{0pt}\rotatebox{0}{\resizebox{.3\textwidth}{!}{\includegraphics{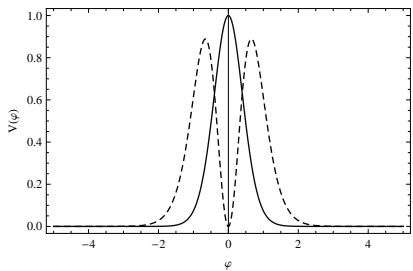}}}
\vspace{0pt}{\caption{The thick line and the dashed line represent the potential $(\alpha_1,\alpha_2)=(0,1)$ and $(\alpha_1,\alpha_2)=(1,-1)$  respectively.} \label{potentials}}
\end{figure}

Motivating by the interesting form of  potential \eqref{pot4}, obtained by explicitly  requiring the Hojman symmetry, we consider now some features  of the related cosmological  model in the phase-space view \cite{cao}. The scalar field potential \eqref{pot4}, can be recast in the suitable form
\begin{equation}\label{pot5}
V(\varphi)=V0 \left[\alpha_{1}\textrm{sech}^4(\alpha\varphi)+\alpha_{2}\textrm{sech}^6(\alpha\varphi)\right]
\end{equation}
We consider the dynamics of this model in a spatially flat FRW universe with Hubble parameter $H$, in which both matter and radiation are present. The evolution equations are
\begin{equation}\label{eve}
\begin{split}
&H^2=\frac{1}{3}(\rho_m +\rho_r + \rho_{\varphi})\\&
\dot{H}=-\frac{1}{2}(\rho_m +\frac{4}{3}\rho_r+\dot{\varphi}^2 )\\&
\ddot{\varphi}+3H\dot{\varphi}+\frac{dV}{d\varphi}=0
\end{split}
\end{equation}
where $\rho_m$ and $\rho_r$ are the matter and radiation energy densities respectively;  $\rho_{\varphi}=\frac{1}{2}\dot{\varphi}^2+V(\varphi)$ is the scalar field energy density. Now, let us define the following dimensionless variables  \cite{cao,copeland}

\begin{equation}\label{defs}
\begin{split}
& x=\frac{1}{\sqrt{6}}\frac{\dot{\varphi}}{H}~~~~~~~~~y=\frac{1}{\sqrt{3}}\frac{\sqrt{V}}{H}\\&
z=\frac{1}{\sqrt{3}}\frac{\sqrt{\rho_r}}{H}~~~~~~~~~\lambda=-\frac{V_{,\varphi}}{ V}
\end{split}
\end{equation}
Using these variables and also evolution equations \eqref{eve}, one can set the following autonomous system

\begin{equation}\label{auto}
\begin{split}
& x'=\sqrt{\frac{3}{2}}\lambda y^2+\frac{3}{2}x(-1+x^2+\frac{1}{3}z^2-y^2) \\&
y'=-\sqrt{\frac{3}{2}}\lambda x y+\frac{3}{2}y (1+x^2+\frac{1}{3}z^2-y^2)\\&
z'=-\frac{z}{2}(1-3x^2-z^2+3y^2)\\&
\lambda'=-\sqrt{6}(\Gamma-1)x
\end{split}
\end{equation}
where prime denotes derivative with respect to $\ln a$, and
\begin{equation}
\Gamma=\frac{V_{,\varphi\varphi}{V}}{V_{,\varphi}^2}
\end{equation}
Moreover, using the definitions \eqref{defs}, one can easily verify the algebraic expressions $\Omega_{\varphi}=x^2+y^2$, $\Omega_{r}=z^2$ and also
\begin{equation}\label{cons}
x^2+y^2+z^2=(1-\Omega_m)\leq 1
\end{equation}
\begin{equation}
\omega_{\varphi}=\frac{P_{\varphi}}{\rho_{\varphi}}=\frac{x^2-y^2}{x^2+y^2}
\end{equation}
where $\Omega_m$, $\Omega_r$ and $\Omega_{\varphi}$ are dimensionless cosmological density parameters ($\Omega_i=\frac{\rho_i}{3H^2}$). Using the dynamical system approach, we will consider two simple cases $(\alpha_1,\alpha_2)=(0,1)$ and $(\alpha_1,\alpha_2)=(1,-1)$. These potentials are shown schematically in Fig \ref{potentials}. It is clear, from the autonomous equations \eqref{auto}, that the phase space is four dimensional ($x$,$y$,$z$,$\lambda$). In order to describe the dynamics via more illustrative figures, we will use the projection of the trajectories onto the $x-y$ plane. Therefore taking into account the bound \eqref{cons}, the dynamics can be described by trajectories within a unit circle in the $x-y$ plane \cite{cao}. Fixed points in the phase space correspond to solutions where the scalar field has a constant equation of state $\omega_{\varphi}$ and the scale factor of the universe evolves as $a\sim t^{\frac{2}{3(1+\omega_{\varphi})}}$. Thus $\omega_{\varphi}<-\frac{1}{3}$ gives rise to an accelerated expansion.
\subsubsection*{$a)~~~V(\varphi)=V_0 \textrm{sech}^6(\alpha\varphi)$}

For this potential the fixed (or critical) points, i.e. the points that verify $x'=y'=z'=\lambda'=0$ are listed in Table \ref{table1}. Depending on the value of $\alpha$ there are only four stable (or attractor) points. If $\alpha^2\geq \frac{1}{12}$ then $A_{\pm}$ and $B_{\pm}$ are attractor points and for $\alpha^2< \frac{1}{12}$, $C_{\pm}$ and $D_{\pm}$ are the attractor points. 


\begin{table*}[!ht]
\begin{center}
\caption{The critical points of $V\sim \textrm{sech}^6(\alpha\varphi)$}
\begin{tabular}{c c c c c c c c c}
\hline\hline
Point &~~~~ $x$ &~~~~ $y$ &~~~~ $z$ &~~~~$\lambda$& ~~~~$(\Omega_r,\Omega _{\phi })$  &~~~~ $\omega_{\varphi}$& ~~~~Existence &~~~~ Stability \\ \hline
$A_{\pm}$ &~~~~ $-\frac{1}{2\sqrt{6}\alpha}$ & ~~~~$\pm\frac{1}{2\sqrt{6}\alpha}$ & ~~~~$0$
& ~~~~$-6\alpha$ & ~~~~$(0,\frac{1}{12\alpha^2})$&~~~~$0$&~~~~$\alpha^2\geq\frac{1}{12}$&~~~~$\alpha^2\geq\frac{1}{12}$  \\

$B_{\pm}$  &~~~~ $\frac{1}{2\sqrt{6}\alpha}$ & ~~~~$\pm\frac{1}{2\sqrt{6}\alpha}$ & ~~~~$0$
& ~~~~$6\alpha$ & ~~~~$(0,\frac{1}{12\alpha^2})$&~~~~$0$&~~~~$\alpha^2\geq\frac{1}{12}$&~~~~$\alpha^2\geq\frac{1}{12}$  \\

$C_{\pm}$ &~~~~ $-\sqrt{6}\alpha$ &~~~~ $\pm\sqrt{1-6\alpha^2}$ &~~~~ $0$ &~~~~ $-6\alpha$&~~~~$(0,1)$&~~~~$(12\alpha^2-1)$&~~~~$\alpha^2\leq\frac{1}{6}$&~~~~$\alpha^2<\frac{1}{12}$\\

$D_{\pm}$  &~~~~ $\sqrt{6}\alpha$ &~~~~ $\pm\sqrt{1-6\alpha^2}$ &~~~~ $0$ &~~~~ $6\alpha$&~~~~$(0,1)$&~~~~$(12\alpha^2-1)$&~~~~$\alpha^2\leq\frac{1}{6}$&~~~~$\alpha^2<\frac{1}{12}$\\

$E$  &~~~~ $0$ &~~~~ $0$ &~~~~ $0$ &~~~~ $\forall \lambda$&~~~~$(0,0)$&~~~~undefined&~~~~$\forall~\alpha,\lambda$&~~~~unstable\\
$F_{\pm}$  &~~~~ $0$ &~~~~ $0$ &~~~~ $\pm 1$ &~~~~ $\forall \lambda$&~~~~$(1,0)$&~~~~undefined&~~~~$\forall~\alpha,\lambda$&~~~~unstable\\
$G_{\pm}$  &~~~~ $0$ &~~~~ $\pm 1$ &~~~~ $0$ &~~~~ $0$&~~~~$(0,1)$&~~~~$-1$&~~~~$\forall~\alpha$&~~~~unstable\\
$H_{\pm}$  &~~~~ $\pm 1$ &~~~~ $0$ &~~~~ $0$ &~~~~ $6\alpha$&~~~~$(0,0)$&~~~~$1$&~~~~$\forall~\alpha$&~~~~unstable\\
$I_{\pm}$  &~~~~ $\pm 1$ &~~~~ $0$ &~~~~ $0$ &~~~~ $-6\alpha$&~~~~$(0,0)$&~~~~$1$&~~~~$\forall~\alpha$&~~~~unstable\\
$J_{\pm}$  &~~~~ $\pm 1$ &~~~~ $0$ &~~~~ $0$ &~~~~ $6\alpha$&~~~~$(0,0)$&~~~~$1$&~~~~$\forall~\alpha$&~~~~unstable\\
$K_{\pm}$  &~~~~ $-\frac{\sqrt{2}}{3\sqrt{3}\alpha}$ &~~~~ $\pm \frac{1}{3\sqrt{3}\alpha}$ &~~~~ $\pm\sqrt{1-\frac{1}{9\alpha^2}}$ &~~~~ $-6\alpha$&~~~~$(1-\frac{1}{9\alpha^2},\frac{1}{9\alpha^2})$&~~~~$\frac{1}{3}$&~~~~$\alpha^2\geq \frac{1}{9}$&~~~~unstable\\
$L_{\pm}$  &~~~~ $\frac{\sqrt{2}}{3\sqrt{3}\alpha}$ &~~~~ $\pm \frac{1}{3\sqrt{3}\alpha}$ &~~~~ $\pm\sqrt{1-\frac{1}{9\alpha^2}}$ &~~~~ $6\alpha$&~~~~$(1-\frac{1}{9\alpha^2},\frac{1}{9\alpha^2})$&~~~~$\frac{1}{3}$&~~~~$\alpha^2\geq \frac{1}{9}$&~~~~unstable\\
 \hline\hline
\label{table1}
\end{tabular}
\end{center}
\end{table*}
%

Let us first consider the unstable points. For every value of $\alpha$, there are six fixed points $H_{\pm}$, $I_{\pm}$ and $J_{\pm}$ that correspond to solutions where the constraint \eqref{cons} is dominated by the kinetic energy of the scalar field. These solutions are unstable, and since the scalar field rolls down the potential energy in the presence of the "friction force" $-3H\dot{\varphi}$ (see equation \eqref{eve}), we expect them to be relevant at early times. On the other hand, the fixed point $G_{\pm}$ correspond to the scalar field dominated solutions ($\Omega_{\varphi}=1$) where the constraint \eqref{cons} is dominated by the potential energy density of the scalar field. For these points $x=\lambda=0$ and so $\dot{\varphi}=\frac{dV}{d\varphi}=0$. This means that $G_{\pm}$ correspond to solutions where the scalar field is frozen at the pick of the potential. It is natural that this points are unstable.

There are also two types of fluid dominated solutions $E$ and $F_{\pm}$. Note that since the coordinate $\lambda$ can take every value in the four dimensional phase space, these solutions represent infinite number of fixed points. $E$ correspond to a matter dominated universe. On the other hand, solutions $F_{\pm}$ correspond to a radiation dominated universe. The fact that $E$ and $F_{\pm}$ are unstable is cosmologically plausible. Because, we do not expect that the universe remains in a fluid dominated phase forever.  Also, this means that the energy density of the scalar field never vanishes with respect to the matter and radiation in the universe. We will show that depending on the value of $\alpha$, the energy density of the scalar field either dominates the other matter fields  in the universe or remains proportional to the matter energy density.

The critical points $K_{\pm}$ and $L_{\pm}$ correspond to the scaling solutions where $\frac{\rho_{\varphi}}{\rho_r}=\frac{1}{9\alpha^2-1}$. The matter energy density for these solutions is zero. Since these solutions are unstable, we can say that the matter energy density never vanishes while both $\rho_{\varphi}$ and $\rho_r$ are nonzero.

Now, let us consider the attractor points. Attractor points $A_{\pm}, B_{\pm}$ differ in several features from $C_{\pm}, D_{\pm}$, so we consider them separately.

\emph{Attractors of type} $A_{\pm}$ \emph{and} $B_{\pm}$. These attractors exist if $\alpha^2\geq \frac{1}{12}$. They correspond to the late time scaling solutions where the energy density of the scalar field remains proportional to that of the matter fluid. Note that for these solutions $\Omega_{r}=0$, and neither the scalar field nor the matter fluid entirely dominates the evolution of the universe. Also the deceleration parameter ($q=-\frac{\ddot{a} }{a H^2}$) for these solutions is $q=\frac{1}{2}$. Therefore, these solutions correspond to a decelerating expansion.

\begin{figure}[!htb]
{\resizebox{.3\textwidth}{!}{\includegraphics{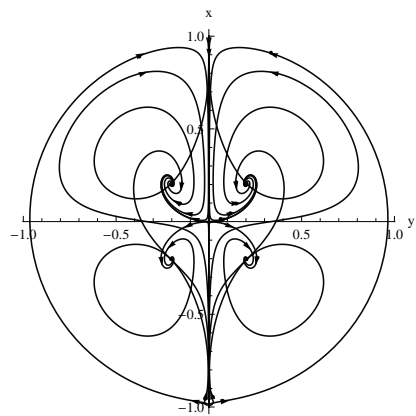}}}
\vspace{0pt}{\caption{Projection of the phase space onto the $x-y$ plane for $\alpha=1$ for the potential $V\sim \textrm{sech}^6(\alpha\varphi)$. The late time attractors are scaling solution with $(x,y)=(\pm\frac{1}{2\sqrt{6}},\pm\frac{1}{2\sqrt{6}})$.} \label{ps1}}
\end{figure}

\begin{figure}[!htb]
\hspace{0pt}\rotatebox{0}{\resizebox{.3\textwidth}{!}{\includegraphics{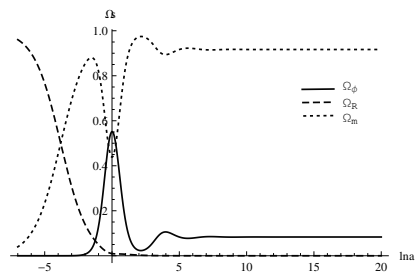}}}
\vspace{0pt}{\caption{Behavior of cosmological density parameters $\Omega_{\varphi}$ (thick line), $\Omega_{r}$ (dashed line) and $\Omega_{m}$ (dotted line) for $\alpha=1$ for the potential $V\sim \textrm{sech}^6(\alpha\varphi)$ as a function of $\ln a$.} \label{O1}}
\end{figure}

\begin{figure}[!htb]
\hspace{0pt}\rotatebox{0}{\resizebox{.3\textwidth}{!}{\includegraphics{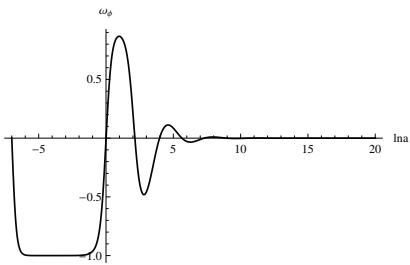}}}
\vspace{0pt}{\caption{The evolution of $\omega_{\varphi}$ for $\alpha=1$ for the potential $V\sim \textrm{sech}^6(\alpha\varphi)$. } \label{o1}}
\end{figure}

%

In Fig. \ref{ps1}, we show the projection of the phase space trajectories onto the $x-y$ plane for $\alpha=1$. In this case, the four attractor points are $(x,y)=(\pm\frac{1}{2\sqrt{6}},\pm\frac{1}{2\sqrt{6}})$. For a proper choice of initial conditions, the trend of $\Omega_m$, $\Omega_r$ and $\Omega_{\varphi}$ is reported in Fig \ref{O1}. Also the evolution of the equation of state parameter $\omega_{\varphi}$ with respect to $\ln a$ has been shown in Fig. \ref{o1}.

\begin{figure}[!htb]
\hspace{0pt}\rotatebox{0}{\resizebox{.3\textwidth}{!}{\includegraphics{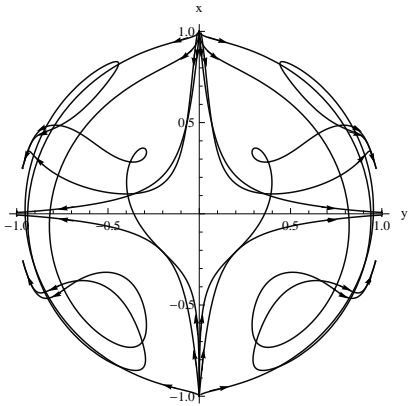}}}
\vspace{0pt}{\caption{Projection of the phase space onto the $x-y$ plane for $\alpha=0.1$ for the potential $V\sim \textrm{sech}^6(\alpha\varphi)$. The late time attractors are scalar field dominated universe with $(x,y)=(\pm 0.245,\pm 0.969)$} \label{ps2}}
\end{figure}

\emph{Attractors of type} $C_{\pm}$ \emph{and} $D_{\pm}$. These attractors exist if $\alpha^2<\frac{1}{12}$. They correspond to the late time scalar field dominated universe ($\Omega_{\varphi}$=1) where the equation of state parameter $\omega_{\varphi}=(12\alpha^2-1)< 0$ and the deceleration parameter $q=(18\alpha^2-1)$. Therefore, the expansion can be accelerated if $\alpha^2<\frac{1}{18}$. The two dimensional subspace of the phase space for $\alpha=0.1$ is reported in Fig. \ref{ps2}. A cosmologically acceptable trajectory should start with the radiation dominated era where $z\sim1$, then enter to the matter dominated era ($z\sim x\sim y\sim0$) and finally fall into one the attractors $C_{\pm}$,  $D_{\pm}$.


\begin{figure}[!htb]
\hspace{0pt}\rotatebox{0}{\resizebox{.3\textwidth}{!}{\includegraphics{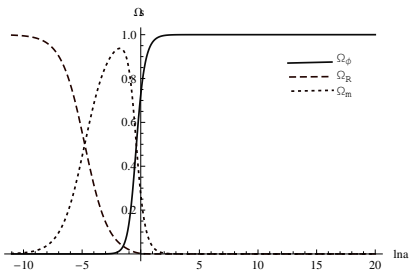}}}
\vspace{0pt}{\caption{The evolution of the density parameters for $\alpha=0.1$ for the potential $V\sim \textrm{sech}^6(\alpha\varphi)$. The initial conditions at $lna=-7$ are $(x,y,z,\lambda)=(10^{-5},1.6\times10^{-5},0.95,1)$.} \label{O2}}
\end{figure}

\begin{figure}[!htb]
\hspace{0pt}\rotatebox{0}{\resizebox{.3\textwidth}{!}{\includegraphics{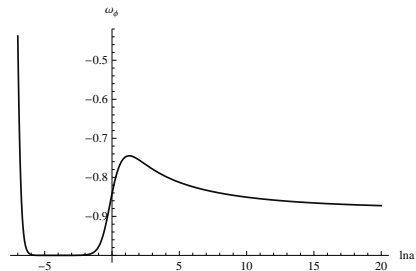}}}
\vspace{0pt}{\caption{Behavior of $\omega_{\varphi}$ for $\alpha=0.1$ for the potential $V\sim \textrm{sech}^6(\alpha\varphi)$. The initial conditions at $lna=-7$ are $(x,y,z,\lambda)=(10^{-5},1.6\times10^{-5},0.95,1)$.} \label{o2}}
\end{figure}
%

In Fig. \ref{O2}, we have chosen a set of initial conditions which leads to a cosmologically acceptable trajectory. However, it is quite clear that in order to be consistent with the observed current matter content of the universe, we have to choice the initial conditions in such a way that the late time scalar field dominated attractor is not yet reached at the present epoch. Therefore, we inevitably confront a kind of fine tuning problem. It is necessary to note that for the potential under consideration the dimensionless parameter $\Gamma<1$, and so it is expected that this model has not tracking solutions \cite{tracking}. In other words, the solutions are quite sensitive to the initial conditions. 

In this case, the evolution of $\omega_{\varphi}$ has been shown in Fig. \ref{o2}. When the solution reaches the attractor, $\omega_{\varphi}$ reaches to the constant value $-0.88$.

\subsubsection*{$b)~~~V(\varphi)=V_0 [\textrm{sech}^4(\alpha\varphi)-\textrm{sech}^6(\alpha\varphi)]$}
In this case, the region of existence and stability of the critical points are listed in Table \ref{table2}. We shall see that the cosmological behavior of this model is similar to the previous model. Depending on the value of $\alpha$, there are only two attractor points. If $\alpha^2\geq \frac{3}{16}$ then there exist two attractor points $M_{\pm}$. Otherwise, if $\alpha^2< \frac{3}{16}$, $N_{\pm}$ are the attractor points.

As in the previous subsection, let us first consider the unstable points. For every value of $\alpha$, there are two critical points $T_{\pm}$ that correspond to solutions where the constraint \eqref{cons} is dominated by the kinetic energy of the scalar field. These solutions are unstable and as we mentioned before, they are expected to be relevant at early times.

The unstable fixed points $S_{\pm}$ correspond to the scalar field dominated solutions ($\Omega_{\varphi}=1$) where the constraint \eqref{cons} is dominated by the potential energy density of the scalar field. Since $\lambda=0$, so these points correspond to solutions where the scalar field is at rest at the pick of the potential.

There are also two types of fluid dominated unstable solutions $R$ and $Q_{\pm}$. $R$ corresponds to a matter dominated universe. $Q_{+}$ and $Q_{-}$ correspond to a expanding radiation dominated universe and to a contracting radiation dominated universe respectively. As we already mentioned, the existence of these critical points means that the energy density of the scalar field never vanishes with respect to the other matter content of the universe.

The critical points $O_{\pm}$ and $P_{\pm}$ correspond to the scaling solutions for which $\frac{\rho_{\varphi}}{\rho_r}=\frac{1}{4\alpha^2-1}$. The matter energy density for these solutions is zero. Since these solutions are unstable, we conclude that the matter energy density never vanishes while $\frac{\rho_{\varphi}}{\rho_r}$ is a nonzero constant.

Now let us consider the attractor points in more detail.

\emph{Attractors of type} $M_{\pm}$. They correspond to the late time decelerating scaling solutions where the scalar field energy density and the matter share a finite and constant portion of the cosmic energy content. $M_{+}$ corresponds to a expanding universe and $N_{-}$ to a contracting universe. For these solutions the radiation energy density is zero and neither the scalar field nor the matter fluid entirely dominates the evolution.

\begin{table*}[!ht]
\begin{center}
\caption{The critical points of $V\sim [\textrm{sech}^4(\alpha\varphi)-\textrm{sech}^6(\alpha\varphi)]$}
\begin{tabular}{c c c c c c c c c}
\hline\hline
Point &~~~~ $x$ &~~~~ $y$ &~~~~ $z$ &~~~~$\lambda$& ~~~~$(\Omega_r,\Omega _{\phi })$  &~~~~ $\omega_{\varphi}$& ~~~~Existence &~~~~ Stability \\ \hline
$M_{\pm}$ &~~~~ $\frac{\sqrt{3}}{\sqrt{32}\alpha}$ & ~~~~$\pm\frac{\sqrt{3}}{\sqrt{32}\alpha}$ & ~~~~$0$
& ~~~~$4\alpha$ & ~~~~$(0,\frac{3}{16\alpha^2})$&~~~~$0$&~~~~$\alpha^2\geq\frac{3}{16}$&~~~~$\alpha^2\geq\frac{3}{16}$  \\

$N_{\pm}$  &~~~~ $\sqrt{\frac{8}{3}}\alpha$ & ~~~~$\pm\sqrt{1-\frac{8}{3}\alpha^2}$ & ~~~~$0$
& ~~~~$4\alpha$ & ~~~~$(0,1)$&~~~~$\frac{16}{3}\alpha^2-1$&~~~~$\alpha^2\leq\frac{3}{8}$&~~~~$\alpha^2<\frac{3}{16}$  \\

$O_{\pm}$  &~~~~ $\frac{1}{\sqrt{6}\alpha}$ &~~~~ $\pm\frac{1}{\sqrt{12}\alpha}$ &~~~~ $-\sqrt{1-\frac{1}{4\alpha^2}}$ &~~~~ $4\alpha$&~~~~$(1-\frac{1}{4\alpha^2},\frac{1}{4\alpha^2})$&~~~~$\frac{1}{3}$&~~~~$\alpha^2\geq\frac{1}{4}$&~~~~unstable\\

$P_{\pm}$  &~~~~ $\frac{1}{\sqrt{6}\alpha}$ &~~~~ $\pm\frac{1}{\sqrt{12}\alpha}$ &~~~~ $\sqrt{1-\frac{1}{4\alpha^2}}$ &~~~~ $4\alpha$&~~~~$(1-\frac{1}{4\alpha^2},\frac{1}{4\alpha^2})$&~~~~$\frac{1}{3}$&~~~~$\alpha^2\geq\frac{1}{4}$&~~~~unstable\\

$Q_{\pm}$  &~~~~ $0$ &~~~~ $0$ &~~~~ $\pm 1$ &~~~~ $\forall \lambda$&~~~~$(1,0)$&~~~~undefined&~~~~$\forall~\alpha$&~~~~unstable\\

$R$  &~~~~ $0$ &~~~~ $0$ &~~~~ $0$ &~~~~ $\forall \lambda$&~~~~$(0,0)$&~~~~undefined&~~~~$\forall~\alpha$&~~~~unstable\\

$S_{\pm}$  &~~~~ $0$ &~~~~ $\pm 1$ &~~~~ $0$ &~~~~ $0$&~~~~$(0,1)$&~~~~$-1$&~~~~$\forall~\alpha$&~~~~unstable\\

$T_{\pm}$  &~~~~ $\pm 1$ &~~~~ $0$ &~~~~ $0$ &~~~~ $0$&~~~~$(0,1)$&~~~~$1$&~~~~$\forall~\alpha$&~~~~unstable\\
 \hline\hline
\label{table2}
\end{tabular}
\end{center}
\end{table*}
%

\begin{figure}[!htb]
\hspace{0pt}\rotatebox{0}{\resizebox{.3\textwidth}{!}{\includegraphics{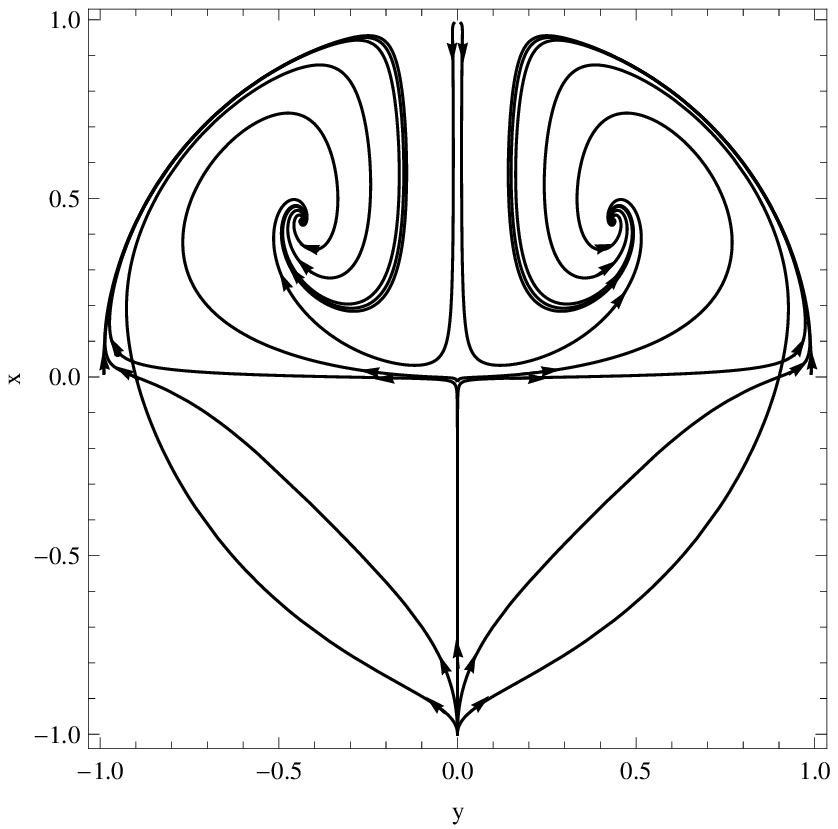}}}
\vspace{0pt}{\caption{Projection of the phase space onto the $x-y$ plane for $\alpha=0.7$. The late time attractors are the scaling solution with $(x,y)=(0.44,\pm0.44)$.} \label{ps3}}
\end{figure}

\begin{figure}[!htb]
\hspace{0pt}\rotatebox{0}{\resizebox{.3\textwidth}{!}{\includegraphics{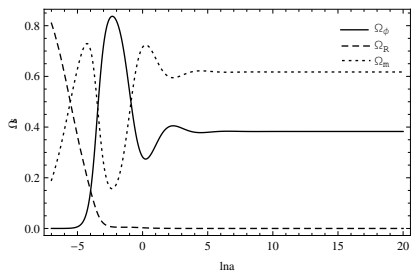}}}
\vspace{0pt}{\caption{Behavior of the cosmological density parameters with respect to $\ln a$ for $\alpha=0.7$.} \label{O3}}
\end{figure}

\begin{figure}[!htb]
\hspace{0pt}\rotatebox{0}{\resizebox{.3\textwidth}{!}{\includegraphics{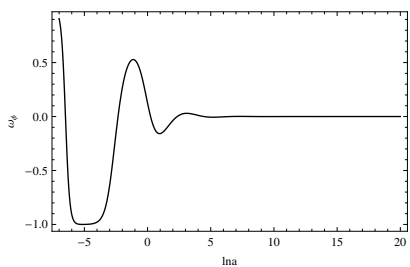}}}
\vspace{0pt}{\caption{Evolution of $\omega_{\varphi}$ with respect to $\ln a$ for $\alpha=0.7$} \label{o3}}
\end{figure}

%

In Fig. \ref{ps3}, we show the projection of the phase space trajectories onto the $x-y$ plane for $\alpha=0.7$. In this case, the late time attractor points are $(x,y)=(0.44,\pm 0.44)$. For a proper choice of initial conditions, the trend of $\Omega_m$, $\Omega_r$ and $\Omega_{\varphi}$ is reported in Fig \ref{O3}. Also, the evolution of the equation of state parameter $\omega_{\varphi}$ with respect to $\ln a$ has been shown in Fig. \ref{o3}.

\begin{figure}[!htb]
\hspace{0pt}\rotatebox{0}{\resizebox{.3\textwidth}{!}{\includegraphics{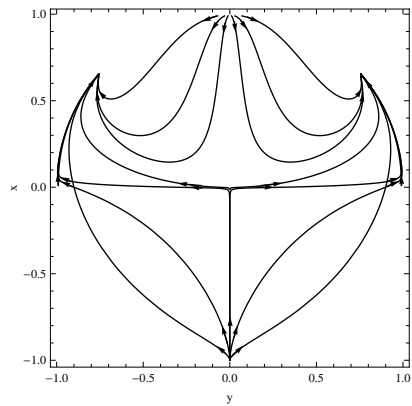}}}
\vspace{0pt}{\caption{Projection of the phase space onto the $x-y$ plane for $\alpha=0.4$. The late time attractors are the scaling solution with $(x,y)=(0.653,\pm 0.757)$.} \label{ps4}}
\end{figure}

\emph{Attractors of type} $N_{\pm}$. They correspond to the late time scalar field dominated universe where the equation of state parameter $\omega_{\varphi}=(\frac{16}{3}\alpha^2-1)< 0$ and the deceleration parameter $q=(-1+8\alpha^2)$. Therefore, the expansion can be accelerated if $\alpha^2<\frac{1}{8}$. The two dimensional subspace of the phase space for $\alpha=0.4$ is shown in Fig. \ref{ps4}.


\begin{figure}[!htb]
\hspace{0pt}\rotatebox{0}{\resizebox{.3\textwidth}{!}{\includegraphics{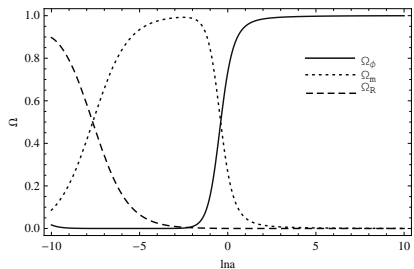}}}
\vspace{0pt}{\caption{Evolution of the cosmological density parameters for $\alpha=0.4$. The initial conditions at $\ln a=-6$ are $(x,y,z,\lambda)=(-10^{-3},2.2\times10^{-4},0.4,1)$.} \label{O4}}
\end{figure}


%

In Fig. \ref{O4}, the behavior of the cosmological density parameters for $\alpha=0.4$ is reported. The initial conditions have been set at $\ln a=-6$ as $(x,y,z,\lambda)=(-10^{-3},2.2\times10^{-4},0.4,1)$. These initial conditions lead to $\Omega_{\varphi}=0.734$ and $\Omega_{m}=0.267$ at the present epoch . In this case, $\alpha^2>\frac{1}{8}$ and so the attractor point corresponds to a decelerating expanding universe.



\begin{figure}[!htb]
\hspace{0pt}\rotatebox{0}{\resizebox{.3\textwidth}{!}{\includegraphics{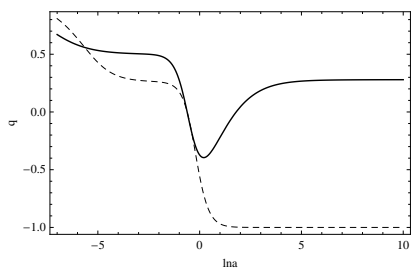}}}
\vspace{0pt}{\caption{Behavior of the deceleration parameter $q$ for $\alpha=0.4$ (thick line) and $q$ for the $\Lambda$CDM model (dashed line). The initial conditions for the $\Lambda$CDM model have been chosen in such a way that lead to $\Omega_{\varphi}=0.734$ and $\Omega_m=0.267$ at present time.} \label{q}}
\end{figure}

%

The behavior of the deceleration parameter $q$ is shown in Fig. \ref{q}. The thick line represents the evolution of the deceleration parameter $q$ with respect to $\ln a$ for $\alpha=0.4$. The dashed line represents $q$ for the $\Lambda$CDM model. As it is clear from this figure, the evolution of the deceleration parameter in the scalar field model is quite different from that of the $\Lambda$CDM model. For both models, $q$ begins at a value near $+1$ in the early times. In both models, at the same time, the universe enters the stage of acceleration. However, in the $\Lambda$CDM model, the universe remains in the accelerated expansion forever. On the other hand, in the case of the scalar field model, the universe stays in the phase of accelerated expansion for a finite time. In other words, the universe will eventually return to the phase of deceleration. When the solution reaches the attractor point, $q$ reaches to the value $q=+0.28$ and the universe stays in the decelerated phase forever. Note that in this phase, although the scalar field (dark energy) is dominating,  the expansion is decelerating.

\section{Conclusion}\label{concl}
In this paper, we introduced a new approach for finding out exact solutions for cosmological models. It is in some senses  similar to the Noether
 symmetry approach but there is no need to construct point-like  Lagrangians generating  the equations of motion.  This means that  the so-called   {\it Hojman symmetries} can exist also for dynamical system where a Lagrangian representation is not possible. 
 
 In particular, we have taken into account scalar-field cosmology models showing how the Hojman symmetry  fixes the scalar field potential and allows the exact integration of dynamics.  Specifically, using this approach we found a class of general interesting  potentials, i.e.  $V(\phi)=V_0\left(\alpha_1\textrm{sech}^4(\alpha\varphi)+\alpha_2\textrm{sech}^6(\alpha\varphi)\right)$ which are useful for dark-energy purposes. We found out exact solutions for the scale factor $a(t)$ and the scalar field $\phi(t)$.
 
 Finally, by a  dynamical system analysis, we investigated this potential for the  two  cases $\alpha_1=0$ and $\alpha_2=-\alpha_1$. By a phase-space representation, we showed that, for both cases, interesting attractor solutions exists. A class of attractors corresponds to the late time scaling solutions where the energy density of the scalar field remains proportional to that of the matter energy density. Another class of attractor solutions corresponds to the late time scalar field dominated universe. Clearly, a detailed analysis in view of observations is out of the scope of the present paper but will be the argument of forthcoming studies.


\end{document}